\documentclass{mem}
\usepackage{natbib}\usepackage{txfonts}\usepackage{balance}
\usepackage{graphicx}
\usepackage[a4paper]{hyperref}
\idline{75}{282}
\begin{document}
\def\teff{$T\rm_{eff }$}
\def\kms{$\mathrm {km s}^{-1}$}

\title{
Accounting for Convective Blue-Shifts in the Determination of 
Absolute Stellar Radial Velocities
}

   \subtitle{}

\author{
C. \,Allende Prieto\inst{1},
L. Koesterke\inst{2},
I. Ram\'{\i}rez\inst{3},
H.-G. Ludwig\inst{4}
\and M. \, Asplund\inst{3}
          }

  \offprints{C. Allende Prieto}

\institute{
Mullard Space Science Laboratory
University College London, UK
\and
Texas Advanced Computing Center,
The University of Texas at Austin, USA
\and
Max Plank Institute for Astrophysics, Garching, Germany
\and
CIFIST GEPI, Observatoire de Paris, CNRS, Universit\'e Paris Diderot, France \\
\email{callende@astro.as.utexas.edu}
}

\authorrunning{Allende Prieto et al.}

\titlerunning{Convective Blue-Shifts}

\abstract{
For late-type non-active stars, gravitational redshifts and convective
blueshifts are the main source of biases in the determination of
radial velocities. If ignored, these effects can introduce systematic errors of
the order of $\sim 0.5$ km s$^{-1}$. We demonstrate that three-dimensional 
hydrodynamical simulations of solar surface convection can be used to
predict the convective blue-shifts of weak spectral lines in solar-like 
stars to $\sim 0.070$ km s$^{-1}$. Using accurate trigonometric parallaxes
and stellar evolution models, the gravitational redshifts can be constrained
with a similar uncertainty, leading to absolute radial velocities accurate 
to $\sim 0.1$ km s$^{-1}$.

}
\maketitle{}

\section{From spectral line Doppler shifts to absolute radial velocities}

A number of factors get in our way when trying to derive absolute
radial velocities of stars from observed wavelengths of spectral lines.
In special relativity, the Lorentz factor affects the observed Doppler
shifts, making it dependent on the velocity component perpendicular
to the line of sight, which is not known a priory. Photons also
become redder as they climb out of the gravitational potential of 
stars; an effect that links
the photon's wavelength to the stellar mass-to-radius ratio.
Pressure shifts, induced by 
collisions between the emitter atom or molecule and other perturbers, 
can also be confused with Doppler shifts. Orbiting companions, 
planets, and activity cycles are other potential sources 
of variability in the observed wavelengths.
In addition, one needs
to disentangle the motion of a star's center of mass from other
motions in its envelope and atmosphere related, for example, to
convection, pulsation, or rotation. 

Recognizing these and other issues involved in the interpretation of
observed Doppler shifts of spectral lines, the IAU has recommended
that research papers report the 'barycentric radial-velocity measure'
(see Lindegren \& Dravins 2003). This is a well-defined quantity that
can be readily determined from spectroscopic observations alone, and which
only includes corrections for the motion of the observer 
relative to the barycenter of the solar system.

Despite a long list of potential distortions,
it is important to underline that for most stars the majority of these
effects amount only to $<0.1$ km s$^{-1}$. {\it Most stars} are 
slowly-rotating non-active late-type dwarfs, and for them 
only two contributors are expected to be important
at the level of several tenths of a kilometer per second: surface
convection and gravitational redshifts. 

The gravitational redshift from the solar photosphere to 1 AU, 
amounts to 633.5 m s$^{-1}$, and the 
same effect on the Earth's gravitational field, reduces
it slightly to 633.3\footnote{On the surface of the Earth, 
the gravitational redshift  along a height of 45m induces
a change of 1$/mu$m s$^{-1}$, and was measured decades ago
using M\"ossbauer spectroscopy (Pound \& Rebka 1960).} m s$^{-1}$.
The gravitational shifts increase slightly for more massive stars
along the main sequence, but are of course much smaller for
giants, and much larger for white dwarfs (see Dravins et al. 1999).
To predict the gravitational redshift, one needs an estimate of
the mass-to-radius ratio of the emitting star. When accurate 
trigonometric parallaxes are available, such as those from Hipparcos
for nearby ($<100$ pc) stars, one can use stellar evolution models
to determine masses and radii of stars within 8\% and 5\%, respectively
(Allende Prieto \& Lambert 1999). For a late-type star, this implies
a 10\% uncertainty in the estimated gravitational redshift, or about
50 m s$^{-1}$.

Surface convection induces offsets in the wavelengths of solar 
spectral lines of up to $\sim$0.5 km s$^{-1}$. This is the net effect
of the upward motion of hot gas (granules) combined with  
narrower and faster downflows (intergranular lanes). Larger
blue-shifts are observed for the spectral lines formed deeper
in the photosphere, where convection strengthens (see, e.g.,
Dravins 1982). 

The stronger lines of atomic iron observed
in the solar spectrum, with an equivalent width larger
than about 200 m\AA, appear free from  net convective shifts 
(Allende Prieto \& Garc\'{\i}a L\'opez 1998). This fact can be
used to empirically correct for such shifts: about two
doze lines measured in the solar atlas of Kurucz et al. (1984) 
give an average of 
623$\pm 12$ m s$^{-1}$. Similar patterns are observed 
in cooler dwarfs, but they are hard to detect in warmer 
objects due to the relative shortage of strong lines
 (see Ram\'{\i}rez et al. in these proceedings).

Three-dimensional hydrodynamical models
offer another route to predict and correct convective
shifts. Asplund et al. (2000a) found an rms scatter of
about 50 m s$^{-1}$ between the predicted and observed 
convective shifts for some 30 Fe I solar lines weaker than 60 m\AA. We
expand that work to include an order of magnitude more lines,
and evaluate a second solar simulation.

\section{Hydrodynamical simulations and spectral synthesis}

In this work we analyze two independent simulations of solar
surface convection from Asplund et al. (2000a,b; hereafter model N) 
and Wedemeyer et al. (2004; hereafter model G). 
Both simulations cover approximately 
a period of 50 minutes of solar time. We used one hundred $50\times50\times82$
snapshots from model N and nineteen $47\times47\times112$ snapshots from
model G to compute the average emergent spectra.

The spectral synthesis was performed with the code ASS$\epsilon$T
(Koesterke et al. 2008). The mean intensity is computed with
short-characteristics and the emergent intensity with long-characteristics.
The flux is calculated from 21 rays. Cubic interpolation is used to
derive the opacities from a pre-calculated grid in $\log T$ (steps
of about 250 K) and $\log \rho$ (steps of 0.25 dex). The equation of
state includes the most important molecules and
ionization stages for elements up to einsteinium ($Z=99$). The 
continuous opacities are derived from the Opacity Project and Iron
Project photoionization cross-sections (see, e.g., Seaton 2005, 
Nahar \& Pradhan 2005). Line opacities are from Kurucz's compilation,
available from his web site, upgraded with Van der Waals damping constants
from Barklem et al. (2000), when available. 

\section{Consistency checks}

It should be emphasized that our choices of equation of
state and opacities for the spectral synthesis calculation are
not consistent with those used in the hydrodynamical simulations. 
Inconsistencies of this kind can vary the location of the
line formation region and artificially modify the predicted
line shifts. 

\begin{figure}[]
\resizebox{\hsize}{!}{\includegraphics[clip=true]{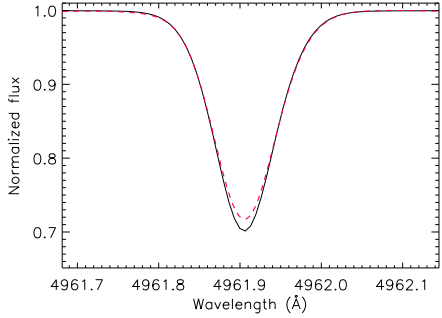}}
\caption{
\footnotesize
Profiles for Fe I $\lambda 4961$ computed with model N and two
different 3D synthesis codes: {\tt lte.x} (solid), and ASS$\epsilon$T
(broken line).
}
\label{f1}
\end{figure}

\begin{table}
\caption{Predicted Fe I convective blue-shifts for model N and two
different synthesis codes}
\label{t1}
\begin{center}
\begin{tabular}{lccc}
\hline
\\

Transition & Residual      & lte.x shift & ASS$\epsilon$T  \\
$~~~~\lambda$ (\AA)  &   flux        & (km s$^{-1}$) &  (km s$^{-1}$)\\
\hline
\\
      4918.9940	& 0.15	& $+0.351$	&	$+0.295$  \\
      4961.9130 & 0.70	& $-0.459$	&   $-0.451$	\\	
      5044.2110	& 0.36	& $-0.251$	&   $-0.271$	\\
      5006.1190	& 0.17	& $+0.279$	& 	$+0.215$  \\
      5049.8200	& 0.19	& $+0.112$	&	$+0.060$  \\
      5068.7660 & 0.23	& $+0.010$	&	$-0.027$	\\
      5076.2640 & 0.39	& $-0.276$	&   $-0.289$	\\
      5096.9980 & 0.30	& $-0.133$	&   $-0.161$	\\
      5098.6980 & 0.24	& $-0.046$	&   $-0.085$	\\
\\
\hline
\end{tabular}
\end{center}
\end{table}

We compared the spectral
synthesis for 9 Fe I lines spanning a significant range of equivalent
widths using model N and both ASS$\epsilon$T, and the spectral
synthesis code {\tt lte.x} (described  by Asplund et al. 2000a and
references therein). While the line profiles show an imperfect 
agreement at a level of a few percent (see the example in 
Fig. \ref{f1}), the predicted
convective shifts, shown in Table \ref{f1}, 
exhibit an rms scatter of 18 m s$^{-1}$ between the two codes, convincing
us that systematic differences in the equation of
state or opacities were small enough to have a very limited impact 
on the predicted line shifts. Note that both codes predict 
positive shifts, i.e. red-shifts, for strong Fe I lines. 

In addition, we checked that when using model N to compute the solar
spectrum in a limited spectral window between 490 and
510 nm, our closest match of the solar atlas was found with an iron abundance 
consistent with that derived by 
Asplund et al. (2000b). Model G would tend to a higher value.

\section{Observed vs. predicted line shifts}

We have systematically measured Fe I line shifts in spectra computed
with model N and model G, as well as in the solar atlas of Kurucz et al. (1984), 
 in the range 400-800 nm.
The models were smoothed to a resolving power of about 430,000 (approximately
that of the solar atlas), 
and with a rotation profile for $v_{\rm rot} \sin i = 1.88$
km s$^{-1}$. The central wavelengths of lines are determined by
fitting third-order polynomials on seven data points
around the line minima. We then subtract the wavelength shifts
measured in the solar atlas from those measured in the model spectra.

\begin{figure*}[]
\resizebox{\hsize}{!}{\includegraphics[clip=true]{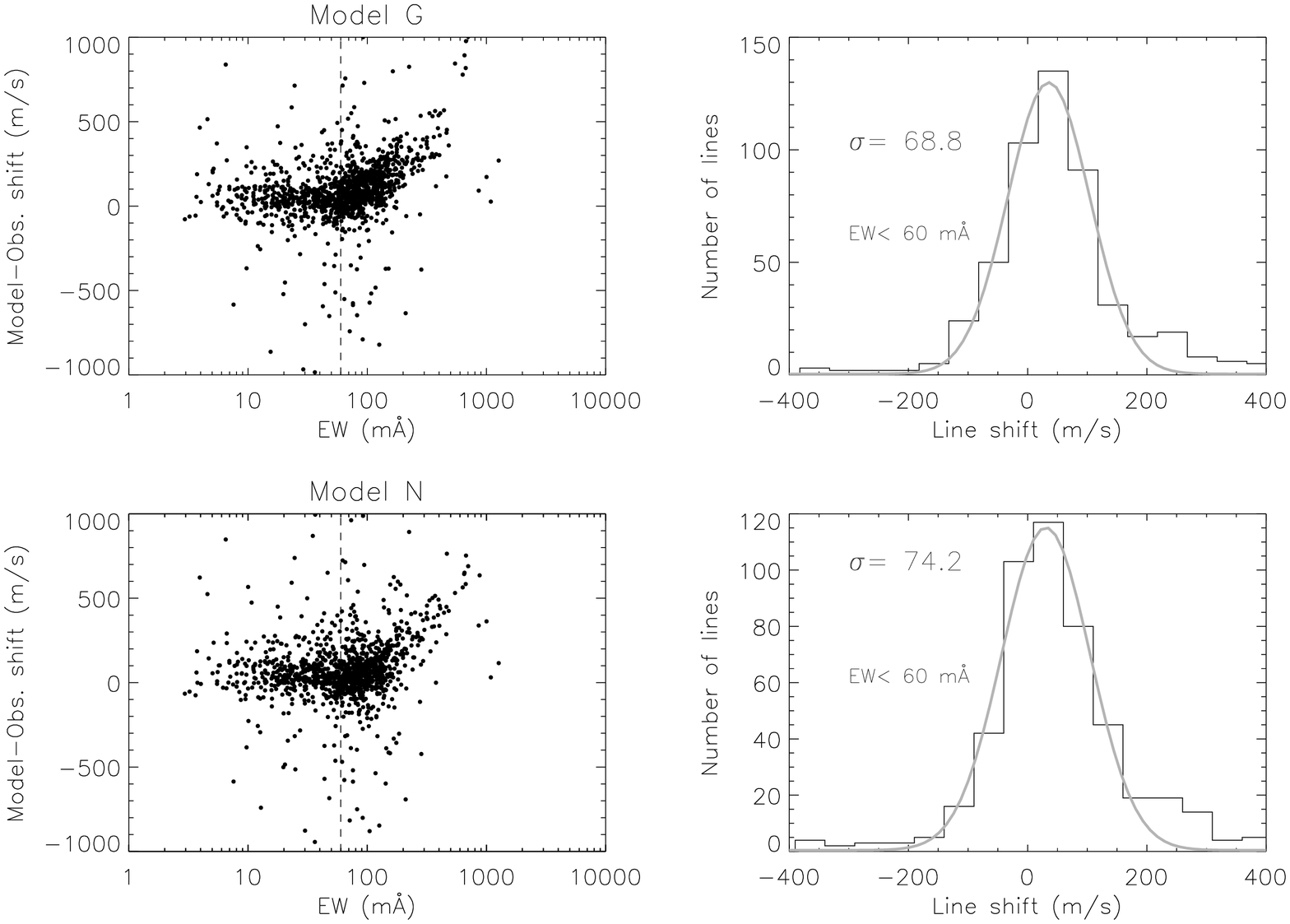}}
\caption{
\footnotesize
Differences between the predicted and the observed Fe I line shifts,
 after subtracting the gravitational shift from the observed
 solar wavelengths. The right-hand panels show histograms
 of the residuals restricted to lines weaker than 60 m\AA.
}
\label{f2}
\end{figure*}

The residuals for both models are displayed in Fig. \ref{f2}.
In agreement with the results of Asplund et al. (2000a), we find
that the predictions from model N are excellent for weak lines,
but the model predicts smaller blueshifts than observed for
lines with equivalent widths stronger than about 60 m\AA. Very
similar results are found for model G. Both models predict
convective redshifts for Fe I lines stronger than about 200 m\AA,
where the observed lines exhibit no convective shifts at all.

By restricting our analysis to lines weaker than 60 m\AA, we
examine the residuals and find that their distributions 
can be approximated by Gaussians with 
a $\sigma$ of about 70 m s$^{-1}$, as shown in Fig. \ref{f2}. 
Both distributions are
slightly biased, indicating that the predicted convective
blue-shifts are slightly smaller in the models than observed
in the atlas by 30--35 m s$^{-1}$. Note that these errors
are not fully associated with imperfections in the models, but
there are uncertainties in the laboratory wavelenghts used,
as well as distorting blends which are not accurately predicted.

\section{Application to Gaia RVS}

At the end of mission, the Radial Velocity Spectrometer 
onboard Gaia will provide radial velocities for some $10^8$ stars 
down to $V\sim17$ mag (Wilkinson et al. 2005). The median error
in the radial velocities will be about 10--15 km s$^{-1}$, however,
results for bright individual stars, or 
averages for stellar systems or populations 
can be much more precise, and should avoid biases due to gravitational
 or convective shifts. 
 
We have calculated the Gaia spectral window (847--874 nm) for
the two solar models discussed above and cross-correlated them,
as well as the observed spectrum, 
with a solar spectrum computed from a static 1D Kurucz model 
(e.g. Castelli \& Kurucz 2003). 
These tests are carried out at a resolving power
much higher than that of the RVS, as at R$\sim11500$ achieving
a precision better than 100 m s$^{-1}$ is challenging.
We find that using a 1D model leads to a systematic error 
of $-263 \pm 3$ m s$^{-1}$. Using any of the 3D models the correct
zero point is found; the cross-correlation between 3D and 1D models
gives offsets of $-258$ and $-279$ m s$^{-1}$ 
for models G and N, respectively.

\section{Conclusions}

Provided with accurate parallaxes, it is possible  
to constrain the mass-to-radius ratio of a late-type star 
to typically 10 \%, and gravitational redshifts to 
$\sim 0.050$  km s$^{-1}$. By computing hydrodynamical
simulations of surface convection, it is also possible to accurately
predict the convective blue-shifts of weak absorption lines to 
$\sim 0.070$ km s$^{-1}$. Therefore, we find that for most non-active late-type stars 
one can predict the biases in the spectroscopic radial velocities associated with
gravitational gravitational and
convective wavelength shifts to within $\sim 0.1$ km s$^{-1}$.

In the case of Gaia mission, which plans to use cross-correlation templates 
calculated from one-dimensional static model atmospheres, 
3D modeling can be used to correct the RVS radial velocities
of late-type stars from convective shifts to within 
a few tens of m s$^{-1}$.

\begin{acknowledgements}
CAP is grateful to the IAU, a Sociedade Astron\^omica Brasileira, 
and UCL graduate school for their generous support. Thanks go 
also to Katia Cunha for her warm hospitality and to Vivien Reuter for 
answering inquiries swiftly and with a smile. 
HGL acknowledges support from EU contract MEXT-CT-2004-014265 (CIFIST).
\end{acknowledgements}

\bibliographystyle{aa}

\end{document}